\documentclass[12pt]{amsart}

\usepackage{graphicx}
\usepackage{amsmath}
\usepackage{amssymb}
\usepackage{amsfonts}
\usepackage{amscd}
\usepackage{epsfig}
\usepackage{verbatim}
\usepackage{algorithm}
\usepackage{algorithmic}
\usepackage{enumerate}
\usepackage{multirow}
\usepackage{rotating}

\newcommand{\ignore}[1]{}
\newcommand{\half}{\frac{1}{2}}

\def \sigmasq{\sigma^2}
\def \sigmasqmu{\sigma^2_{\mu}}
\def \sigmasqi{\Sigma^2_i}
\def \cbar{\bar{c}}
\def \ybar{\bar{y}}
\def \Cbar{\bar{C}}
\def \Ybar{\bar{Y}}
\def \bigcbar{\bar{C}}
\def \bigybar{\bar{Y}}
\def \sumi{\sum^n_{i=1}}
\def \sint{S_{int}}
\def \sext{S_{ext}}

\def \sinttilde{\tilde{S}_{int}}
\def \sexttilde{\tilde{S}_{ext}}

\def \var{{\text{Var}}}
\def \cov{{\text{Cov}}}
\def \var{{\text{Var}}}
\def \cov{{\text{Cov}}}

\def \boldbigc{{\bf C}}
\def \boldbigy{{\bf Y}}

\begin{document}

\title [Estimating intrinsic and extrinsic noise]{Estimating intrinsic and extrinsic noise from single-cell gene expression measurements}
\author{Audrey Qiuyan Fu}
\address{Department of Genetics, Stanford University; Department of Human Genetics, University of Chicago; Current Address: Department of Statistical Science, University of Idaho}
\author{Lior Pachter}
\address{Departments of Mathematics, Molecular \& Cell Biology and Computer Science, UC Berkeley}
\begin{abstract}
Gene expression is stochastic and displays variation (``noise'') both within and
between cells. Intracellular (intrinsic) variance can be distinguished
from extracellular (extrinsic) variance by applying the law of total variance to data from two-reporter assays
that probe expression of identical gene pairs in single-cells. We examine established formulas for the 
estimation of intrinsic and extrinsic noise and provide interpretations of them in terms of a hierarchical model. This allows us 
to derive corrections that minimize the mean squared error, an objective that may be important when sample sizes are small.  The statistical 
framework also highlights the need for quantile normalization, and provides justification for the use of the sample correlation 
between the two reporter expression levels to estimate the percent contribution of extrinsic noise to the total noise. Finally, we provide a geometric interpretation of these results that clarifies the current interpretation. 

\end{abstract}
\maketitle

\section{Introduction}
In a classic paper on the stochasticity of gene expression, Elowitz
{\em et al.} \cite{Elowitz2002} describe a clever two-reporter
expression assay designed to tease apart ``intrinsic'' and
``extrinsic'' noise from the overall variability in gene
expression. The idea is as follows: two identically regulated reporter genes (cyan fluorescent protein and yellow fluorescent protein) are inserted into individual {\it E. coli.} cells allowing for comparable expression measurements within and between cells.
 If $n$ cells are assayed, this leads to expression
measurements $c_1,\ldots,c_n$ and $y_1,\ldots,y_n$ where the pair
$(c_i,y_i)$ represent the expression measurements for the cyan and yellow
reporters in the $i$th cell. The goal of the experiment is to measure
the variance in gene expression from the pairs $(c_i,y_i)$ (denoted by
$\eta^2_{tot}$) and to
ascribe it to two different sources: first, variability due to the different states of cells (``extrinsic noise'', denoted by
$\eta^2_{ext}$), and second, inherent variability that exists even
when the state of cells is fixed (``intrinsic noise'', denoted by
$\eta^2_{int}$). In \cite{Elowitz2002}, formulas were provided for estimating $\eta^2_{ext}, \eta^2_{int}$
and $\eta^2_{tot}$ (hereafter referred to as the ELSS estimates) that were later interpreted in terms of the ``law of total variance'' in \cite{Hilfinger2011}:

\begin{eqnarray}
\eta^2_{int} & = & \frac{\frac{1}{n} \left( \sum_{i=1}^n \frac{1}{2} (c_i-y_i)^2 \right)} {\overline{c} \cdot \overline{y}},  \label{eqn:noise.int.elowitz} \\
\eta^2_{ext} & = & \frac{\frac{1}{n}\sum_{i=1}^n c_i \cdot y_i -
\overline{c} \cdot \overline{y}}{\overline{c} \cdot \overline{y}},  \label{eqn:noise.ext.elowitz}\\
\eta^2_{tot} & = & \frac{ \frac{1}{n}\sum_{i=1}^n
\frac{1}{2}(c_i^2+y_i^2) -  \overline{c} \cdot \overline{y}}{\overline{c} \cdot \overline{y}},  \label{eqn:noise.tot.elowitz}
\end{eqnarray}
where $\overline{c} = \frac{1}{n}\sum_{i=1}^n c_i$ and $\overline{y} =
\frac{1}{n}\sum_{i=1}^n y_i$. 

\section{A hierarchical model}
\label{sec:model}
Although the work of \cite{Hilfinger2011} sheds light on the statistical basis of the ELSS estimators, it does not address questions about their statistical properties, such as bias and accuracy. To analyze these aspects of the estimators we introduce a hierarchical model that provides a formal model for the experiments of \cite{Elowitz2002}.

In the rest of the paper, we focus on the numerators of (\ref{eqn:noise.int.elowitz},\ref{eqn:noise.ext.elowitz},\ref{eqn:noise.tot.elowitz}). They are the key components of the formulas and can be viewed as estimators of true variances. We note that lower case letters such as $c_i$ and $y_i$ denote observations not only in the ELSS formulas but throughout our paper; we reserve uppercase letters for random variables.  

A hierarchical model for expression of the two reporters in a cell emerges naturally from the assumption that reporter expression, conditioned on the same cellular environment, is represented by independent and identically distributed random variables. To allow each cell to be different from the others, we introduce independent identically distributed random variables $Z_i,$ for $i=1,\ldots,n$ that represent the environments of cells (as in \cite{Hilfinger2011}). We posit that the cellular conditional random variables associated to the two reporters have the same distribution $F$ with mean $M_i$ and variance 
$\sigma^2_i$,  both parameters being unique to the $i$-th cell:
\begin{align}
C_i | Z_i &\sim F(M_i, \sigmasqi) \mbox{ and}  \label{eqn:model.ci}\\
Y_i | Z_i &\sim F(M_i, \sigmasqi). 	\label{eqn:model.yi}
\end{align}
Thinking of a two reporter experiment as ``random'', in the sense that the states of cells $Z_1,\ldots,Z_n$ are random, across cells we have
\begin{align*}
M_i &\sim G(\mu, \sigmasqmu) \mbox{ and}\\
\sigmasqi &\sim H(\sigmasq, \epsilon),
\end{align*}
where $G$ is the distribution of all the $M_i$s, with mean $\mu$ and variance $\sigmasqmu$, and $H$ that of all the 
$\sigmasqi$s, with mean $\sigma^2$ and variance $\epsilon$.  In other words, both the mean and variance of reporter expression level is cell specific and the random variable $\sigmasqi$ and its mean $\sigmasq$
represent the ``within-cell" variation as distinguished from the parameter $\sigmasqmu$ which represents the ``between-cell" variability in the 
ANOVA setting.

For any $i$, the mean of $C_i$ or $Y_i$ is $\mu$, according to the following calculation:
\begin{align}
E[C_i] = E_{Z_i}[E[C_i|Z_i]] = E[M_i] = \mu. \label{eqn:meanlaw}
\end{align}
The total variance in $C_i$ (or $Y_i$) can be calculated using the ``law of total variance'':
\begin{equation}
Var[C_i] = E_{Z_i}[Var[C_i|Z_i]] + Var_{Z_i}[E[C_i|Z_i]].   \label{eqn:tvlaw}
\end{equation}
Using the notation of the hierarchical model described above, and dropping the subscripts for expectation because they are clear by context, we have, for any $i$,
\begin{align}
E[Var[C_i|Z_i]] &= \sigmasq \;\;\;\; & \text{(within-cell variability; intrinsic noise)}, \label{eq:intrnoise}\\
Var[E[C_i|Z_i]] &= \sigmasqmu \;\;\;\; & \text{(between-cell variability; extrinsic noise)} \label{eq:extnoise}.
\end{align}
With this notation equation (\ref{eqn:tvlaw}) becomes
\begin{align}
Var[C_i] = E[Var[C_i|Z_i]] + Var[E[C_i|Z_i]] = \sigmasq + \sigmasqmu \;\;\;\;\;\;\;\;\; & \text{(total noise)} \label{eq:totnoise}.
\end{align}
This means that the marginal (unconditional) distributions of $C_i$ and $Y_i$ are identical:
\begin{align*}
C_i &\sim N (\mu, \sigmasq + \sigmasqmu); \\
Y_i &\sim N (\mu, \sigmasq + \sigmasqmu).
\end{align*}

In the next sections, we will derive the estimators for intrinsic and extrinsic noise, and examine the bias and mean squared error (MSE)
of each estimator.  
Specifically, for any estimator $S$, the MSE of $S$ with respect to the true parameter $\tau$ is calculated as follows:
\begin{align*}
E [(S - \tau)^2] &= E [S - E[S] + E[S] - \tau]^2 \\
		      &= E \bigg[ (S-E[S])^2 + (E[S]-\tau)^2 + 2(S-E[S])(E[S]-\tau) \bigg] \\
		      &= E[S-E[S]]^2 + E[E[S]-\tau]^2  \\
		      &= Var[S] + (E[S] - \tau)^2,
\end{align*}
where $E[S]-\tau$ is the bias of $S$.

\section{Intrinsic noise}
\label{sec:intrinsic}
Starting with the law of total variance, the within-cell variability
$E[Var[C_i|Z_i]]$ for cell $i$ can be written as:
\begin{align}
E[Var[C_i|Z_i]] &= Var[C_i] - Var[E[C_i|Z_i]] \notag\\
&= \half [Var[C_i] + Var[Y_i]] - Cov[C_i,Y_i] \notag\\
&= \half [Var[C_i] - 2Cov[C_i,Y_i] + Var[Y_i]] \notag\\
&= \half Var[C_i-Y_i]. \notag\\
& = \half \left( E[C_i-Y_i]^2 - (E[C_i-Y_i])^2\right) \notag
\end{align}


This leads to the following unbiased estimator for the intrinsic noise:
\begin{align*}
S^*_{int} &= \frac{1}{2(n-1)} \sumi \bigg[(C_i - Y_i) - (\Cbar - \Ybar)\bigg]^2 \\
			  &= \frac{1}{2(n-1)} \sumi (C_i - Y_i)^2 - \frac{n}{2(n-1)} (\Cbar - \Ybar)^2.
\end{align*}

To find the estimator that minimizes the MSE, we consider estimators of the following general form
\begin{align}
\sint = \frac{1}{2a} \left( \sum^n_1 (C_i - Y_i)^2 - n(\Cbar - \Ybar)^2 \right).  \label{eqn:intrinsic.general}
\end{align}
Assuming normality of the distribution $G$ (i.e., cell-specific means $M_i$ follow a normal distribution), as well as $\mu=0$ and $\epsilon=0$, 
the MSE is given by
\begin{align*}
E [\sint - \sigmasq]^2 &= Var[ \sint] + (E [\sint] - \sigmasq)^2 \\
	&= \frac{1}{2a^2} \bigg[ (2n^2 + \frac{6}{n} - 7)\sigma^4 + 2(\frac{2}{n}-1)\sigmasq \sigmasqmu + \frac{1}{n} \sigma^4_\mu \bigg]  - 2(n-1)\sigma^4 \frac{1}{a} + \sigma^4. 
\end{align*}
The value of $a$ that minimizes this expression is
\begin{align*}
a &= \frac{(2n^3 - 7n + 6) \sigma^4 + 2(2-n)\sigmasq \sigmasqmu + \sigma^4_\mu}{2(n^2 - n) \sigma^4} \\
	&= \frac{2n^3 - 7n + 6}{2(n^2 - n)} + \frac{2-n}{n^2-n} \frac{\sigmasqmu}{\sigmasq} + \frac{1}{2(n^2-n)} \bigg( \frac{\sigmasqmu}{\sigmasq} \bigg)^2.
\end{align*}
See Appendices \ref{sec:app.moments} and \ref{sec:app.mse.intrinsic} for the complete derivation.

The analysis above can be simplified with an additional assumption, namely that $\Cbar = \Ybar$. In some experiments this may be a natural assumption to make, whereas in others the condition is likely to be violated; we comment on this in more detail in the discussion. Here we proceed to note that assuming that $\Cbar = \Ybar$, the estimator (\ref{eqn:intrinsic.general}) simplifies to
\begin{align*}
\sinttilde &= \frac{1}{2a} \sumi (C_i - Y_i)^2.
\end{align*}
The unbiased estimator with this form is easily derived by observing that
\begin{align*}
E[\sinttilde] &= \frac{1}{2a} \sumi E[C_i - Y_i]^2  = \frac{1}{2a} \sumi Var [C_i - Y_i] \\
	    &= \frac{n}{2a} (2\sigmasq + 2\sigmasqmu - 2\sigmasqmu) = \frac{n}{a} \sigmasq.
\end{align*}
Thus, in order for $\sinttilde$ to be unbiased the parameter $a$ must be equal to $n$. The resulting formula is the ELSS formula in (\ref{eqn:noise.int.elowitz}). This makes clear that the assumption $\Cbar = \Ybar$ underlies the derivation of the ELSS intrinsic noise estimator.

In order to study the mean squared error and derive an estimator that minimizes it, we again assume normality of $G$. The MSE of $\sint$ is then given by
\begin{align*}
E[\sinttilde - \sigmasq]^2 &= Var[\sinttilde] + (E[\sinttilde] - \sigmasq)^2 \\
				&= \frac{n}{a^2} (3\epsilon + 2\sigma^4) + (\frac{n}{a} \sigmasq - \sigmasq)^2.
\end{align*}
Assuming again that $\mu=0$ and $\epsilon=0$, the MSE simplifies to
\begin{align*}
E[\sinttilde - \sigmasq]^2 &= \frac{2n}{a^2}\sigma^4 + \sigma^4 \left( \left(\frac{n}{a} \right)^2 - \frac{2n}{a} + 1 \right) \\
							  &= \frac{n\sigma^4(n+2)}{a^2} - \frac{2n\sigma^4}{a} + \sigma^4,
\end{align*}
which is minimized when $a=n+2$ (see Appendices \ref{sec:app.moments} and \ref{sec:app.var.int} for the complete derivation).

\newpage
\section{Extrinsic noise}
\label{sec:extrinsic}

To examine estimators for extrinsic noise, we again start with the law of total variance, this time noting that the within-cell variability
$Var[E[C_i|Z_i]]$ can be written as:

\begin{eqnarray}
Var[E[C_i|Z_i]]   & = &  E[E[C_i|Z_i]^2]-E[E[C_i|Z_i]]^2  \notag\\
& =  & E[E[C_i|Z_i]E[Y_i|Z_i]]-E[E[C_i|Z_i]]^2 \notag\\
& =  & E[E[C_iY_i|Z_i]]-E[E[C_i|Z_i]E[E[Y_i|Z_i]] \notag\\
& = &  E[C_iY_i] - E[C_i]E[Y_i] \notag\\
& = &  Cov[C_i,Y_i].  \label{eqn:cov}
\end{eqnarray}
This connection between the extrinsic noise, the law of total variance and the covariance of $C_i$ and $Y_i$ was noted by Hilfinger and Paulsson in \cite{Hilfinger2011}. 

Formula (\ref{eqn:cov}) leads to the following unbiased estimator for the extrinsic noise, as it is an unbiased estimator estimator for 
the covariance:
\begin{align*}
\sext^* = \frac{1}{n-1} \left( \sumi C_i Y_i - n \Cbar \Ybar\right).
\end{align*}
We note that the ELSS estimator (\ref{eqn:noise.ext.elowitz}) uses the scalar $1/n$, which unlike the case of the intrinsic noise estimator (\ref{eqn:noise.int.elowitz}) leads to a biased estimator in this case.

In order to find the estimator that minimizes the MSE, we consider the following general estimator:
\begin{align*}
\sext = \frac{1}{b}\left( \sumi C_i Y_i - n \Cbar \Ybar\right).
\end{align*}
We again assume that $M_i$ is normal and that $\mu=0$ and $\epsilon=0$. The MSE of $\sext$ is
\begin{align*}
E[\sext - \sigmasqmu]^2 &= \frac{n-1}{b^2} (\sigma^2 + \sigma_\mu^2)^2 + \frac{(n-1)^2}{nb^2} \sigma_\mu^4 + \bigg(\frac{n-1}{b} \sigmasqmu - \sigmasqmu \bigg)^2 \\
	&= (n-1)(\sigma^2 + \sigma_\mu^2)^2\frac{1}{b^2} + (n-1)^2 \bigg( 1+\frac{1}{n} \bigg) \sigma_\mu^4 \frac{1}{b^2} - 2(n-1)\sigma_\mu^4 \frac{1}{b} + \sigma_\mu^4 \\
	&= \left( (n-1)(\sigma^2 + \sigma_\mu^2)^2 + (n-1)^2 \left( 1+\frac{1}{n} \right) \sigma_\mu^4 \right)\frac{1}{b^2} - 2(n-1)\sigma_\mu^4 \frac{1}{b} + \sigma_\mu^4,
\end{align*}
which is minimized when
\begin{align*}
\frac{1}{b} = \frac{\sigma_\mu^4}{(\sigma^2 + \sigma_\mu^2)^2 + (n-1)\bigg( 1+\frac{1}{n} \bigg)\sigma_\mu^4}, \mbox{ or equivalently}
\end{align*}
\begin{align}
b = (n-1)\bigg( 1+\frac{1}{n} \bigg) + \bigg(\frac{\sigma^2 + \sigmasqmu}{\sigmasqmu}\bigg)^2 = (n-1)\bigg( 1+\frac{1}{n} \bigg) + \frac{1}{\rho(\boldbigc,\boldbigy)^2} \label{eq:bforext}.
\end{align}

It is interesting to note that (\ref{eq:bforext}) comprises two parts: the first, $(n-1)(1+\frac{1}{n})$ converges to $n-1$ as $n \rightarrow \infty$, while the second, $(\frac{\sigma^2 + \sigmasqmu}{\sigmasqmu})^2$ is equal to $\frac{1}{\rho(\boldbigc,\boldbigy)^2}$ where $\rho(\boldbigc,\boldbigy)$ is the correlation between vectors $\boldbigc$ and $\boldbigy$. See Appendices \ref{sec:app.moments} and \ref{sec:app.var.ext} for more details.

\section{Geometric interpretation}
\label{sec:geometry}
Figure 3a of \cite{Elowitz2002} shows a scatterplot of data $(c_i,y_i)$ for an experiment and suggests thinking of intrinsic and extrinsic noise geometrically in terms of projection of the points onto a pair of orthogonal lines. While this geometric interpretation of noise agrees exactly with the ELSS intrinsic noise formula, the interpretation of extrinsic noise is more subtle. Here we complete the picture.

\begin{figure}[!ht]
\includegraphics[scale=0.6]{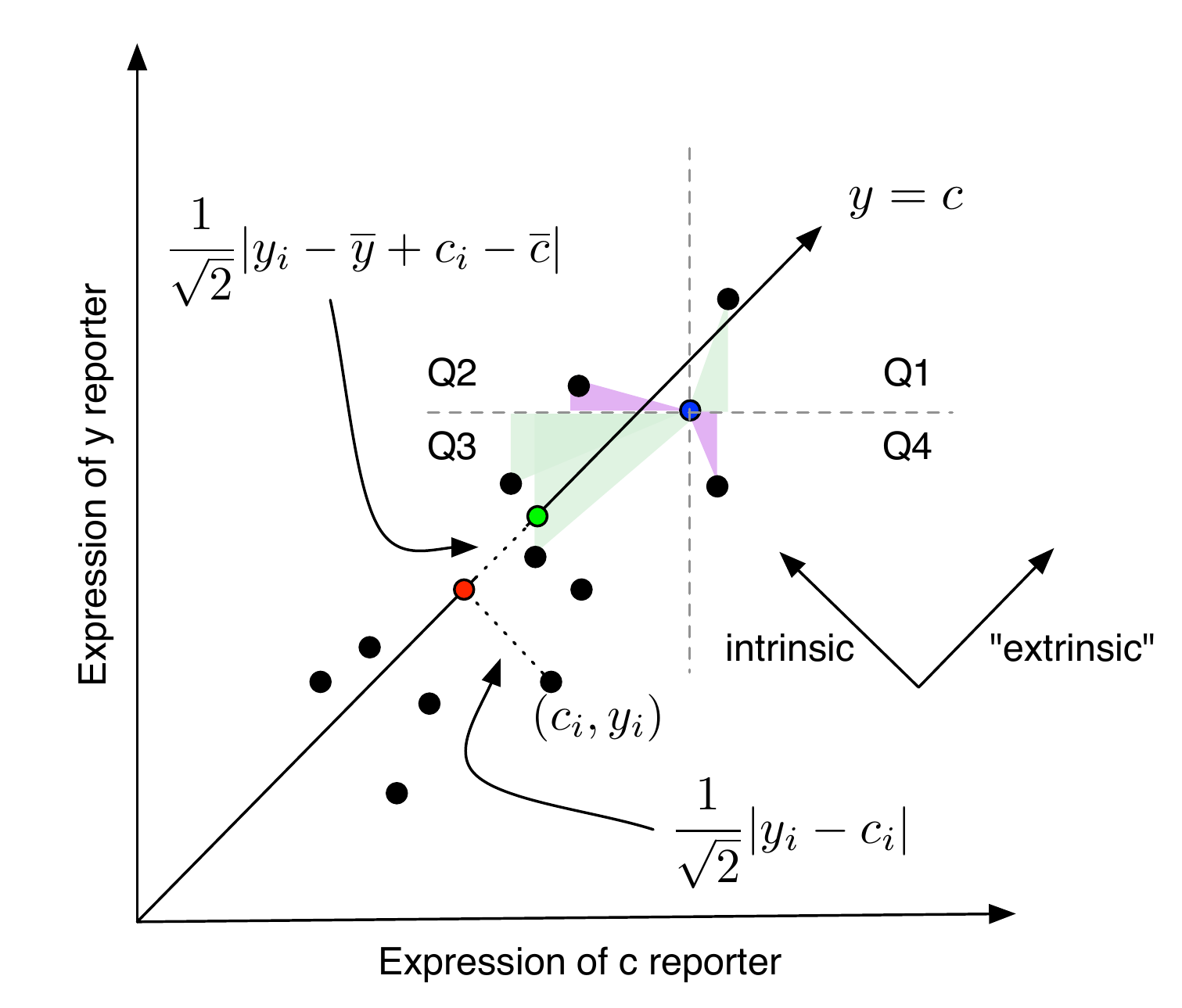}
\caption{Geometric interpretation of intrinsic and extrinsic noise.  The intrinsic noise, or the within-cell variability, 
is the variance of the points projected to the line $y=-c$, which is perpendicular to $y=c$. In other words, it is the average of the squared lengths $\frac{1}{2}(y_i-c_i)^2$.  The red point is the projection of point $(c_i, y_i)$ onto the line $y=c$.  The green point is the centroid.  See the main text for additional detail.  The extrinsic noise, or the between-cell 
variability, is the sample covariance between $c_i$ and $y_i$. 
The colored triangles around the blue point illustrate the geometric interpretation 
of the sample covariance: it is the average (signed) area of triangles formed by pairs of data points: green triangles in Q1 and Q3 (some not shown) represent a positive contribution to the covariance, whereas the magenta triangles in Q2 and Q4 a negative contribution.
Since most data points lie in the 1st (Q1) and 3rd (Q3) quadrants relative to the blue point, most of the contribution involving 
the blue point is positive.  Similarly, since most pairs of data points can be connected by a positively signed line, their positive contribution will result in a positive covariance. In \cite{Elowitz2002} the direction along the line $y=c$ is labeled extrinsic, which makes sense in terms of the intuition for positive sample covariance. However we have placed that label ``extrinsic'' in quotes because the extrinsic noise estimator corresponding directly to the sample variance for points projected onto the line $y=c$ (in analogy with intrinsic noise) is heavily biased and not usable in practice.} 
\label{fig:noise}
\end{figure}

To understand the intuition behind Figure 3a in \cite{Elowitz2002}, we have redrawn it in a format that highlights the math (Fig~\ref{fig:noise}). The projection of a point $(c_i,y_i)$ onto the line $y=c$ is the point $(\frac{1}{2}(y_i+c_i),\frac{1}{2}(y_i + c_i))$, shown as the red point
in Fig.~\ref{fig:noise}. The intrinsic noise, as estimated by the unbiased estimator (\ref{eqn:noise.int.elowitz}) is then the mean squared distance from the origin to the points projected onto the line $y=-c$. 

The ELSS estimate for the extrinsic noise is the sample covariance.  Intuitively, it indicates how the measurements of one reporter track that of the other across cells.  
The geometric meaning of the sample covariance in Fig.~\ref{fig:noise} is based on an alternative formulation of sample covariance \cite{Hayes2011,Heffernan1988}:
\begin{align*}
\cov ({\bf c}, {\bf y}) = \frac{1}{n-1} \sum_{i=1}^n (c_i - \cbar) (y_i - \ybar).
\end{align*}
This formulation of the sample covariance has the interpretation of being an average of the signed area of triangles associated to pairs of points, and is very different from what might be considered at first glance an appropriate anology to intrinsic noise, namely the sample variance along the line $y=c$.

The estimate corresponding to the sample variance of the projected points along the line $y=c$, using as a mean the projected centroid $(\frac{\cbar+\ybar}{2},\frac{\cbar+\ybar}{2})$ which is shown as the green point in Fig.~\ref{fig:noise}, turns out to be biased by an amount equal to the total noise. Using
\begin{align*}
\sexttilde^{*} & = \frac{1}{n-1} \sum_{i=1}^n 2\left(\frac{1}{2}(Y_i-\Ybar+C_i-\Cbar)\right)^2\\
& = \frac{1}{2(n-1)} \sumi \left( (C_i + Y_i)^2 - (\Cbar+\Ybar) ^2 \right)
\end{align*}  
the bias is
\begin{align*}
E[\sexttilde^{*}] - \sigmasqmu &= \frac{1}{2} Var [C_i + Y_i] - \sigmasqmu\\
& = \frac{1}{2} \left(Var[C_i] + Var[Y_i] + 2 Cov [C_i, Y_i] \right) -\sigmasqmu \\
& = \frac{1}{2} \left( 2(\sigmasq + \sigmasqmu) + 2\sigmasqmu\right) -\sigmasqmu = \sigmasq + \sigmasqmu
\end{align*}
which is the true total noise.

The above calculation also shows that if the intrinsic and extrinsic noise are both estimated as variances along the projections to the lines $y=-c$ and $y=c$ respectively, then the total noise will be overestimated by a factor of two. 

In summary, the caption to Figure 3a in \cite{Elowitz2002} is completely accurate in stating that ``Spread of points perpendicular to the diagonal line on which CFP and YFP intensities are equal corresponds to intrinsic noise, whereas spread parallel to this line is increased by extrinsic noise.'' However the geometric interpretation of covariance makes precise exactly {\em how} an increase in extrinsic noise relates to the spread of points in the direction of the line $y=c$. 

\section{Practical considerations}
\subsection{Data normalization}
Our hierarchical model, as well as the ANOVA interpretation, is consistent with the model in Elowitz
{\em et al.} \cite{Elowitz2002}; both models assume that within each
cell there are two distributions for the expression of the two
reporter genes and that they have the same true mean and true variance. With the normality assumption, this means that 
the two reporters have identical distributions.  Elowitz {\em et al.} measured the single-color distributions of strains that 
contained lac-repressible promoter pairs, which verified that this was
a reasonable assumption in the case of cyan fluorescent protein (CFP) and yellow fluorescent protein (YFP) in their experiment.  

Other studies have adapted this system and used other reporter combinations 
that may have markedly different distributions.  For example, 
Yang {\em et al.}  \cite{Yang.etal.2014} used CFP and mCherry with vastly different ranges of intensity values: whereas CFP varied from 0 to 6000 
(arbitrary units; i.e., a.u.), mCherry could vary from 0 to 9000 (a.u.); see 
Fig. 3a from their paper. 
In contrast, another study \cite{Schmiedel2015} normalized the two reporters used in their experiment (ZsGreen and mCherry) to have the same mean.  However, 
the variances, or more generally, 
the two distributions, also need to be the same.  
Since the decomposition of the total noise depends on the assumption that both reporters in the same cellular environment 
have similar variance (see (\ref{eqn:model.ci}) and (\ref{eqn:model.yi})), we 
recommend that in general a quantile normalization which normalizes
the reporter measurements to identical distributions be performed before
the calculations of noise components.  Such a normalization procedure
is standard in many settings requiring similar assumptions. 

\subsection{Optimal estimators for intrinsic and extrinsic noise}
We have derived the estimators 
that are optimal for minimizing bias or the MSE (summarized in Table~\ref{table:estimators}).  The ELSS estimator in (\ref{eqn:noise.int.elowitz})
is in fact a special case of the general estimator under the assumption that $\Cbar=\Ybar$, and is appropriate for data that are normalized 
to have the same sample mean (i.e., $\cbar=\ybar$).  In \cite{Elowitz2002}, the intensities of the two reporters were normalized to have mean 1. In the case where the assumption of equal reporter means does not hold, the general estimator is more suitable.

Similar to the estimators for the intrinsic noise, we derived two estimators for extrinsic noise, optimized for bias and for 
MSE respectively (Table~\ref{table:estimators}). 

The sample size $n$ is the leading term in the denominator of all the optimal (in either the bias or MSE sense) intrinsic and extrinsic noise estimators. As a result, the unbiased estimator has the same form as the min-MSE estimator for large $n$ (Table~\ref{table:estimators}).  
For extrinsic noise, the general estimates converge to the ELSS estimate (Table~\ref{table:estimators}).  For intrinsic noise, assuming $\cbar=\ybar$, 
the ELSS estimate is optimal for bias and MSE for large $n$ and optimal for bias at small $n$.
Indeed, in \cite{Elowitz2002}, typical values for $n$ are greater than 100, making the ELSS formulas suitable for the analyses performed (with the assumption 
of equal mean satisfied). However, our derivations indicate that the two types of noise can be estimated using fewer cells.

As a general rule we recommend computing the inverse squared correlation between the $c_i$ and $y_i$ values and applying a correction if it is comparable (up to a small factor) to the sample size.
\begin{sidewaystable}
\vspace*{14cm}
   \caption{Estimators for intrinsic and extrinsic noise}
\bigskip
\centering
\begin{tabular}{c|p{1.6in}|p{3.0in}|p{2.1in}}
\hline
 & \multicolumn{2}{|c|}{Exact Estimator for Small $n$} &  Large $n$ \\
 \cline{2-3}
 & Minimizing Bias (Unbiased) & Minimizing MSE & \\
 \hline\hline
 {\bf Intrinsic noise} \\\\
 \cline{2-4}
General & $\frac{1}{2(n-1)} \bigg[\sum^n_1 (C_i - Y_i)^2 - n(\bigcbar - \bigybar)^2 \bigg]$ & $\frac{1}{2a} \bigg[\sum^n_1 (C_i - Y_i)^2 - n(\bigcbar - \bigybar)^2 \bigg]$, where $a = \frac{2n^3 - 7n + 6}{2(n^2 - n)} + \frac{2-n}{n^2-n} \frac{\sigmasqmu}{\sigmasq} + \frac{1}{2(n^2-n)}  \bigg( \frac{\sigmasqmu}{\sigmasq} \bigg)^2$ & $\frac{1}{2n} \bigg[\sum^n_1 (C_i - Y_i)^2 - n(\bigcbar - \bigybar)^2 \bigg]$\\
Assuming $\Cbar=\Ybar$ & $\frac{1}{2n} \sumi (C_i - Y_i)^2$ & $\frac{1}{2(n+2)} \sumi (C_i - Y_i)^2$ & $\frac{1}{2n} \sumi (C_i - Y_i)^2$ \\
  & (ELSS estimator) & & (ELSS estimator) \\
\hline
 {\bf Extrinsic noise} \\\\
 \cline{2-4}
 General & $\frac{1}{n-1} (\sumi C_i Y_i - n \bigcbar \bigybar)$ & $\frac{\sigma_\mu^4}{(\sigma^2 + \sigma_\mu^2)^2 + (n-1)\bigg( 1+\frac{1}{n} \bigg)\sigma_\mu^4} \left( \sumi C_i Y_i - n \bigcbar \bigybar \right)$ & $\frac{1}{n} (\sumi C_i Y_i - n \bigcbar \bigybar)$ \\
 & & & (ELSS estimator) \\
 \hline\hline
\end{tabular}
\label{table:estimators}
\end{sidewaystable}
\newpage
\subsection{Assessing the ratio of extrinsic to intrinsic noise from sample correlation}
\label{sec:corr}


We have seen that the proportion of the between-cell variability to total variability is the correlation $\rho(\boldbigc,\boldbigy)$. This leads to a simple approach for estimating the relative magnitude of the two types of noise: one can compute the 
sample correlation of the expression of the two reporters, $\rho({\bf
  c},{\bf y})$, and the ratio of extrinsic to 
intrinsic noise is then estimated by $\rho({\bf c},{\bf y})/[1-\rho({\bf c},{\bf y})]$.  For example, in Elowitz et al \cite{Elowitz2002}, the sample correlation $\rho({\bf c},{\bf y})$ is roughly 0.7, which implies that about 70\% of the total noise is extrinsic noise and the ratio of extrinsic to intrinsic noise is 2.33.

\section{Acknowledgments}

This project began as a result of discussion during a journal club
meeting of Jonathan Pritchard's group that A.F. was attending. We
thank Michael Elowitz and Peter Swain for facilitating reanalysis of the data from \cite{Elowitz2002}. A.F. was partially supported by K99HG007368
(NIH/NHGRI). L.P. was partially supported by NIH grants R01 HG006129
and R01 DK094699.

\newpage
\appendix
\section{Moments of $M_i$ and $C_i$ under normality}
\label{sec:app.moments}
Assuming that $M_i  \sim N(\mu, \sigmasqmu)$, we have
\begin{align*}
E[M_i - \mu]^3 &= 0; \\
E[M_i - \mu]^4 &= 3\sigma_{\mu}^4.
\end{align*}
We can compute the third and fourth moments of $M_i$ as follows:
\begin{align*}
E[M_i - \mu]^3 &= E[M_i^2 + \mu^2 - 2M_i \mu)(M_i - \mu] \\
			   &= E[M_i^3 - 2M_i^2 \mu + M_i \mu^2 - M_i^2 \mu - \mu^3 + 2M_i \mu^2] \\
			   &= E[M_i^3 - 3M_i^2 \mu + 3M_i \mu^2 - \mu^3]  \\
			   &= E[M_i^3] - 3\mu (\sigma^2_\mu + \mu^2) + 3\mu^3 - \mu^3 \\
			   &= E[M_i^3] - 3\mu \sigma^2_\mu - \mu^3,
\end{align*}
which gives
\begin{align*}
E[M_i^3]= 3\mu \sigma^2_\mu + \mu^3.
\end{align*}
\begin{align*}
E[M_i - \mu]^4 &= E[M^2_i - 2M_i \mu + \mu^2]^2  \\
			   &= E[M^4_i + \mu^4 + 4M^2_i \mu^2 + 2M^2_i \mu^2 - 4M^3_i \mu - 4M_i \mu^3]  \\
			   &= E[M^4_i + \mu^4 + 6M^2_i \mu^2 - 4M^3_i \mu - 4M_i \mu^3] \\
			   &= E[M^4_i] + \mu^4 + 6\mu^2 (\sigma^2_\mu + \mu^2) - 4\mu(3\mu \sigma^2_\mu + \mu^3) - 4\mu^4 \\
			   &= E[M^4_i] + \mu^4 + 6\mu^2 \sigma^2_\mu + 6\mu^4 - 12\mu^2 \sigma^2_\mu - 4\mu^4 - 4\mu^4 \\
			   &= E[M^4_i] - 6\mu^2 \sigma^2_\mu - \mu^4,
\end{align*}
which gives
\begin{align*}
E[M^4_i] = 3\sigma^4_\mu + 6\mu^2 \sigma^2_\mu + \mu^4.
\end{align*}
For the random variable $C_i$, since $\sigmasqi \sim H(\sigmasq, \epsilon)$, such that 
\begin{align*}
E[\sigmasqi] &= \sigmasq; \\
\var[\sigmasqi] &= \epsilon,
\end{align*}
we have
\begin{align*}
E[C^4_i] &= E[E[C^4_i| Z_i]] \\
	     &= E[3\Sigma^4_i + 6M^2_i \Sigma^2_i + M^4_i] \\
	     &= 3(\epsilon + \sigma^4) + 6(\sigmasqmu + \mu^2) \sigmasq + 3\sigma^4_\mu + 6\mu^2 \sigma^2_\mu + \mu^4 \\
	     &= 3\epsilon + 3\sigma^4 + 6\sigmasqmu \sigmasq + 6\mu^2 \sigmasq + 3\sigma^4_\mu + 6\mu^2 \sigma^2_\mu + \mu^4.
\end{align*}
Further assuming that $\mu=0$, i.e., the means are all 0, and that $\epsilon=0$, which means that the variability is the same across cells, we have
\begin{align*}
E[M_i^3] &= 0 \\
E[M_i^4] &= 3\sigma^4_\mu;
\end{align*}
and
\begin{align*}
E[C^3_i] &= 0 \\
E[C^4_i] &= 3 (\sigmasq + \sigmasqmu)^2.
\end{align*}

\section{MSE of the general intrinsic noise estimator}
\label{sec:app.mse.intrinsic}
The general form of the estimator for intrinsic noise is
\begin{align*}
S = \frac{1}{2a} \left(\sum^n_1 (C_i - Y_i)^2 - n(\Cbar - \Ybar)^2 \right).
\end{align*}
Thus
\begin{align*}
Var[S] = \frac{1}{4a^2} \left( Var \left[\sum(C_i - Y_i)^2] + n^2 Var[(\bigcbar - \bigybar)^2\right] - 2n Cov \bigg[\sum(C_i - Y_i)^2, (\bigcbar - \bigybar)^2 \bigg] \right).
\end{align*}
Below we will assume normality, as well as $\mu=0$ and $\epsilon=0$, to facilitate the derivation.

First, we note that
\begin{align*}
Var [(\bigcbar - \bigybar)^2] &= Var [\bigcbar^2 - 2\bigcbar \bigybar + \bigybar^2] \\
	&= Var[\bigcbar^2] + 4Var[ \bigcbar \bigybar] + Var [\bigybar^2]- 4 Cov [\bigcbar^2, \bigcbar \bigybar] - 4 Cov [\bigybar^2, \bigcbar \bigybar] + 2 Cov [\bigcbar^2, \bigybar^2].
\end{align*}
\begin{align*}
Var [\bigcbar^2 ]&= Var \bigg[\frac{C_1 + \dots + C_n}{n} \cdot \frac{C_1 + \dots + C_n}{n}\bigg] \\
	&= \frac{1}{n^4} Var \left[\sum C^2_k + \sum_{i\neq j} C_i C_j\right] \\
	&= \frac{1}{n^4} \left( Var \sum C^2_k + Var [\sum_{i \neq j} C_i C_j] + 2 Cov [\sum C^2_k, \sum_{i \neq j} C_i C_j] \right) \\
	&= \frac{1}{n^4} \left( 2n(\sigmasq + \sigmasqmu)^2 + n(n-1)(\sigmasq + \sigmasqmu)^2 + 0 \right) \\
	&= \frac{n+1}{n^3} (\sigmasq + \sigmasqmu)^2.
\end{align*}
This is because
\begin{align*}
Var \left[\sum_{i \neq j} C_i C_j\right] &= \sum_{i \neq j} Var[ C_i C_j] \\
	&= \sum_{i \neq j} \left( EC^2_i C^2_j - (EC_i C_j)^2 \right) \\
	&= \sum_{i \neq j} \left( (\sigmasq + \sigmasqmu)^2 - 0 \right) \\
	&= n(n-1) (\sigmasq + \sigmasqmu)^2.
\end{align*}

Additionally,
\begin{align*}
Var[ \bigcbar \bigybar] &= \frac{1}{n^2} Var [n\bigcbar \bigybar] \\
	&= \frac{1}{n^2} \left( (\sigmasq + \sigmasqmu)^2 + \frac{\sigma^4_\mu}{n} \right) \\
	&= \frac{1}{n^2} (\sigmasq + \sigmasqmu)^2 + \frac{\sigma^4_\mu}{n^3}.
\end{align*}

\begin{align*}
Cov [\bigcbar^2, \bigcbar \bigybar]&= \frac{1}{n^4} Cov \left[\sum C^2_k + \sum_{i \neq j} C_i C_j, \sum C_l Y_l + \sum_{m \neq r} C_m C_r\right] \\
	&= \frac{1}{n^4} \bigg( Cov \left[\sum C^2_k, \sum C_l Y_l\right] + Cov \left[\sum C^2_k, \sum_{m \neq r} C_m C_r\right] \\
	& \;\;\;\; + Cov \left[\sum_{i \neq j} C_i C_j, \sum C_l Y_l\right] + Cov \left[\sum_{i \neq j} C_i C_j, \sum_{m \neq r} C_m C_r\right] \bigg).
\end{align*}
\begin{align*}
Cov \left[\sum C^2_k, \sum C_l Y_l\right] &= Cov \left[\sum C^2_k, \sum C_k Y_k\right] \\
	&= \sum (E[C^3_k Y_k] - E[C^2_k] E[C_k Y_k]) \\
	&= \sum \bigg[ 3\sigmasq \sigmasqmu + 3 \sigma^4_\mu - (\sigmasq + \sigmasqmu) \sigmasqmu \bigg] \\
	&= 2n \sigmasqmu (\sigmasq + \sigmasqmu).
\end{align*}
For $Cov \left[\sum C^2_k, \sum_{m \neq r} C_m C_r\right]$, since
\begin{align*}
Cov [C^2_i, C_i Y_j] = E[C^3_i Y_j] - E[C^2_i] E[C_i Y_j] = 0
\end{align*}
and 
\begin{align*}
Cov [C^2_i, C_j Y_k] = E[C^2_iC_j Y_k] - E[C^2_i] E[C_j Y_k] = 0,
\end{align*}
we have
\begin{align*}
Cov \left[\sum C^2_k, \sum_{m \neq r} C_m C_r\right] = 0.
\end{align*}
For $Cov \left[\sum_{i \neq j} C_i C_j, \sum C_l Y_l\right]$, since
\begin{align*}
Cov [C_i C_j, C_i Y_i] = E [C^2_i Y_i C_j]- E[ C_i C_j] E[ C_i Y_i] = 0
\end{align*}
and 
\begin{align*}
Cov [C_k C_l, C_i Y_i] = E [C_k C_l C_i Y_i ]- E[ C_k C_l] E[ C_i Y_i] = 0,
\end{align*}
we have
\begin{align*}
Cov \left[\sum_{i \neq j} C_i C_j, \sum C_l Y_l\right] = 0.
\end{align*}
Additionally,
\begin{align*}
Cov \left[\sum_{i \neq j} C_i C_j, \sum_{m \neq r} C_m C_r\right] &= \sum_{i, j, m, r} Cov [C_i C_j, C_m C_r] \\
	&= \sum_{i\neq j} Cov [C_i C_j, C_i C_j] \\
	&= \sum_{i \neq j} Var [C_i C_j] \\
	&= n(n-1) (\sigmasq + \sigmasqmu)^2.
\end{align*}
Therefore,
\begin{align*}
Cov [\bigcbar^2, \bigcbar \bigybar] = \frac{2}{n^3} \sigmasqmu (\sigmasq + \sigmasqmu) + \frac{n-1}{n^3} (\sigmasq + \sigmasqmu)^2.
\end{align*}

Furthermore,
\begin{align*}
Cov [\bigcbar^2, \bigybar^2] &= \frac{1}{n^4} Cov \left[\sum C^2_k + \sum_{i \neq j} C_i C_j, \sum Y^2_l + \sum_{m \neq r} Y_m Y_r\right] \\
	&= \frac{1}{n^4} \bigg( Cov \left[\sum C^2_k, \sum Y^2_l\right] + Cov \left[\sum C^2_k, \sum_{m \neq r} Y_m Y_r\right] \\
	& \;\;\;\; + Cov\left[\sum Y^2_l, \sum_{i \neq j} C_i C_j\right] + Cov \left[\sum_{i \neq j} C_i C_j, \sum_{m \neq r} Y_m Y_r\right] \bigg).
\end{align*}
In the expression above,
\begin{align*}
Cov \left[\sum C^2_k, \sum Y^2_l\right] = 2n \sigma^4_\mu;
\end{align*}
\begin{align*}
Cov \left[\sum C^2_k, \sum_{m \neq r} Y_m Y_r\right] =  Cov \left[\sum Y^2_l, \sum_{i \neq j} C_i C_j\right] = 0;
\end{align*}
\begin{align*}
Cov \left[\sum_{i \neq j} C_i C_j, \sum_{m \neq r} Y_m Y_r\right] &= \sum_{i \neq j} Cov [C_i C_j, Y_i Y_j] \\
	&= \sum_{i \neq j} (E[ C_i C_j Y_i Y_j] - E[ C_i C_j ]E[ Y_i Y_j]) \\
	&= \sum_{i \neq j} (E[C_i Y_i] E[ C_j Y_j] - 0) \\
	&= n(n-1) \sigma^4_\mu.
\end{align*}
Then we have
\begin{align*}
Cov [\bigcbar^2, \bigybar^2] &= \frac{1}{n^4} \bigg( 2n \sigma^4_\mu + n(n-1) \sigma^4_\mu \bigg) \\
	&= \frac{n+1}{n^3} \sigma^4_\mu.
\end{align*}

Putting the terms together, we have
\begin{align*}
Var [\bigcbar - \bigybar]^2 &= Var [\bigcbar^2] + 4Var [\bigcbar \bigybar ]+ Var [\bigybar^2]
		 - 4 Cov [\bigcbar^2, \bigcbar \bigybar] - 4 Cov [\bigybar^2, \bigcbar \bigybar] + 2 Cov [\bigcbar^2, \bigybar^2] \\
		 &= \frac{2(n+1)}{n^3} (\sigmasq + \sigmasqmu)^2 + \frac{4}{n^2} (\sigmasq + \sigmasqmu)^2 + \frac{4\sigma^4_\mu}{n^3} - \frac{16}{n^3} \sigmasqmu (\sigmasq + \sigmasqmu) - \frac{8(n-1)}{n^3} (\sigmasq + \sigmasqmu)^2 \\
		 & \;\;\;\; + \frac{2(n+1)}{n^3} \sigma^4_\mu \\
		 &= \frac{2}{n^3} \bigg( (6-n) (\sigmasq + \sigmasqmu)^2 - 8\sigmasqmu (\sigmasq + \sigmasqmu) + (n+3) \sigma^4_\mu \bigg) \\
		 &= \frac{2}{n^3} \left( (6-n)\sigma^4 + (4-2n) \sigmasq \sigmasqmu + \sigma^4_\mu \right).
\end{align*}

Next, we note that
\begin{align*}
  Cov \left[\sum(C_i - Y_i)^2, (\bigcbar - \bigybar)^2 \right] &= \sum Cov \bigg[ (C_i - Y_i)^2, (\bigcbar - \bigybar)^2 \bigg] \\
	&= \sum \bigg( E[(C^2_i - 2 C_i Y_i + Y^2_i)(\bigcbar^2 - 2\bigcbar \bigybar + \bigybar^2)] \\
	& \;\;\;\; - E[(C^2_i - 2 C_i Y_i + Y^2_i)] E[(\bigcbar^2 - 2\bigcbar \bigybar + \bigybar^2)] \bigg),
\end{align*}
where
\begin{align*}
& \;\;\;\; E[(C^2_i - 2 C_i Y_i + Y^2_i) (\bigcbar^2 - 2\bigcbar \bigybar + \bigybar^2)] \\
&= E\bigg[ C^2_i \bigcbar^2 - 2 C_i Y_i \bigcbar^2 + Y^2_i \bigcbar^2 - 2 C^2_i \bigcbar \bigybar + 4 C_i Y_i \bigcbar \bigybar - 2 Y^2_i \bigcbar \bigybar + C^2_i \bigybar^2 - 2 C_i Y_i \bigybar^2 + Y^2_i \bigybar^2 \bigg],
\end{align*}
and
\begin{align*}
E[C^2_i - 2 C_i Y_i + Y^2_i]  = 2(\sigmasq + \sigmasqmu) - 2 \sigmasqmu = 2\sigmasq,
\end{align*}
\begin{align*}
E[\bigcbar^2 - 2\bigcbar \bigybar + \bigybar^2] = \frac{2}{n} (\sigmasq + \sigmasqmu) - \frac{2}{n} \sigmasqmu = \frac{2\sigmasq}{n}.
\end{align*}

\begin{align*}
E [C^2_i \bigcbar^2] &= \frac{1}{n^2} E [C^2_i \left(\sum C^2_k + \sum_{i \neq j} C_i C_j\right) ]\\
	&= \frac{1}{n^2} \bigg( E[C^4_i] + \sum_{k \neq i} E[C^2_i] E[C^2_k] + \sum_{i \neq j} E [C^2_k C_i C_j] \bigg) \\
	&= \frac{1}{n^2} \bigg[ 3(\sigmasq + \sigmasqmu)^2 + (n-1) (\sigmasq + \sigmasqmu)^2 + 0 \bigg] \\
	&= \frac{n+2}{n^2} (\sigmasq + \sigmasqmu)^2.
\end{align*}
\begin{align*}
E[C_i Y_i \bigcbar^2] &= E \bigg[ C_i Y_i \frac{\sum C^2_j + \sum_{k \neq l} C_k C_l}{n^2} \bigg] \\
	&= \frac{1}{n^2} \bigg( E[ C_i Y_i C^2_i ]+ \sum_{j \neq i} E[ C_i Y_i C^2_j ]+ \sum_{k \neq l} E [C_i Y_i C_k C_l ]\bigg) \\
	&= \frac{1}{n^2} \bigg( 3(\sigmasq \sigmasqmu + \sigma^4_\mu) + (n-1)(\sigmasq \sigmasqmu + \sigma^4_\mu) + 0 \bigg) \\
	&= \frac{n+2}{n^2} \sigmasqmu (\sigmasq + \sigmasqmu).
\end{align*}
\begin{align*}
E[Y^2_i \bigcbar^2] &= E \bigg[ Y^2_i \frac{\sum C^2_j + \sum_{k \neq l} C_k C_l}{n^2} \bigg] \\
	&= \frac{1}{n^2} \bigg( E[ Y^2_i C^2_i] + \sum_{j \neq i} E[ Y^2_i C^2_j ]+ \sum_{k \neq l} E[Y^2_i C_k C_l] \bigg) \\
	&= \frac{1}{n^2} \bigg( (\sigmasq + \sigmasqmu)^2 + 2 \sigma^4_\mu + (n-1)(\sigmasq + \sigmasqmu)^2 + 0 \bigg) \\
	&= \frac{1}{n^2} \bigg( n(\sigmasq + \sigmasqmu)^2 + 2\sigma^4_\mu \bigg).
\end{align*}
\begin{align*}
E[C^2_i \bigcbar \bigybar] &= \frac{1}{n^2} \bigg( E[C^2_i C_i Y_i ]+ \sum_{j \neq i} E[ C^2_i C_j Y_j] + \sum_{k \neq l} E [C^2_i C_k Y_l ]\bigg) \\
	&= \frac{1}{n^2} \bigg( 3(\sigmasq \sigmasqmu + \sigma^4_\mu) + (n-1)(\sigmasq \sigmasqmu + \sigma^4_\mu) + 0 \bigg) \\
	&= \frac{n+2}{n^2} \sigmasqmu (\sigmasq + \sigmasqmu).
\end{align*}
\begin{align*}
E[C_i Y_i \bigcbar \bigybar ]&= \frac{1}{n^2} \bigg( E[C^2_i Y^2_i  ]+ \sum_{j \neq i} E[ C_i Y_i C_j Y_j] + \sum_{k \neq l} E [C_i Y_i C_k Y_l] \bigg) \\
	&= \frac{1}{n^2} \bigg( (\sigmasq + \sigmasqmu)^2 + 2\sigma^4_\mu+ (n-1) \sigma^4_\mu + 0 \bigg) \\
	&= \frac{1}{n^2} \bigg( (\sigmasq + \sigmasqmu)^2 + (n+1) \sigma^4_\mu \bigg).
\end{align*}
Additionally,
\begin{align*}
E[Y^2_i \bigcbar \bigybar] &= E[C^2_i \bigcbar \bigybar] = \frac{n+2}{n^2} \sigmasqmu (\sigmasq + \sigmasqmu); \\
E [C^2_i \bigybar^2] &= E[Y^2_i \bigcbar^2] = \frac{1}{n^2} \bigg( n(\sigmasq + \sigmasqmu)^2 + 2\sigma^4_\mu \bigg); \\
E[C_i Y_i \bigybar^2] &= E[C_i Y_i \bigcbar^2] = \frac{n+2}{n^2} \sigmasqmu (\sigmasq + \sigmasqmu); \\
E [Y^2_i \bigybar^2] &= E[ C^2_i \bigcbar^2] = \frac{n+2}{n^2} (\sigmasq + \sigmasqmu)^2.
\end{align*}
Therefore,
\begin{align*}
& \;\;\;\; E[(C^2_i - 2 C_i Y_i + Y^2_i) (\bigcbar^2 - 2\bigcbar \bigybar + \bigybar^2) ]\\
&= E\bigg[C^2_i \bigcbar^2 - 2 C_i Y_i \bigcbar^2 + Y^2_i \bigcbar^2 - 2 C^2_i \bigcbar \bigybar + 4 C_i Y_i \bigcbar \bigybar - 2 Y^2_i \bigcbar \bigybar + C^2_i \bigybar^2 - 2 C_i Y_i \bigybar^2 + Y^2_i \bigybar^2 \bigg] \\
&= \frac{2(n+2)}{n^2} (\sigmasq + \sigmasqmu)^2 - \frac{4(n+2)}{n^2} \sigmasqmu (\sigmasq + \sigmasqmu) + \frac{2}{n^2} \bigg( n(\sigmasq + \sigmasqmu)^2 + 2\sigma^4_\mu \bigg) \\
& \;\;\;\; - \frac{4(n+2)}{n^2} \sigmasqmu (\sigmasq + \sigmasqmu) + \frac{4}{n^2} \bigg( (\sigmasq + \sigmasqmu)^2 + (n+1) \sigma^4_\mu \bigg) \\
&= \frac{4(n+2) \sigma^4}{n^2}.
\end{align*}

So we have
\begin{align*}
Cov \bigg[\sum(C_i - Y_i)^2, (\bigcbar - \bigybar)^2 \bigg] &= \sum Cov \bigg[ (C_i - Y_i)^2, (\bigcbar - \bigybar)^2 \bigg] \\
	&= \sum \bigg( E(C^2_i - 2 C_i Y_i + Y^2_i) (\bigcbar^2 - 2\bigcbar \bigybar + \bigybar^2) \\
	& \;\;\;\; - E(C^2_i - 2 C_i Y_i + Y^2_i) E(\bigcbar^2 - 2\bigcbar \bigybar + \bigybar^2) \bigg) \\
	&= n \bigg( \frac{4(n+2) \sigma^4}{n^2} - 2\sigmasq \frac{2\sigmasq}{n} \bigg) \\
	&= \frac{8 \sigma^4}{n}.
\end{align*}

The variance of the estimator is then
\begin{align*}
Var [S] &= \frac{1}{4a^2} \bigg( Var \left[\sum(C_i - Y_i)^2 \right]+ n^2 Var[\bigcbar - \bigybar]^2 - 2n Cov \bigg[\sum(C_i - Y_i)^2, (\bigcbar - \bigybar)^2 \bigg] \bigg) \\
	&= \frac{1}{4a^2} \bigg( 8n \sigma^4 + \frac{2}{n} \bigg( (6-n)\sigma^4 + (4-2n) \sigmasq \sigmasqmu + \sigma^4_\mu \bigg) - 16 \sigma^4 \bigg) \\
	&= \frac{1}{2a^2} \bigg( 4n \sigma^4 + \frac{1}{n} \bigg( (6-n)\sigma^4 + (4-2n) \sigmasq \sigmasqmu + \sigma^4_\mu \bigg) - 8 \sigma^4 \bigg).
\end{align*}

The expectation of the estimator is
\begin{align*}
E[ S] = \frac{1}{2a} \bigg( \sum E[C_i - Y_i]^2 - n E[\bigcbar - \bigybar]^2 \bigg),
\end{align*}
where
\begin{align*}
E[(C_i - Y_i)^2] &= Var [C_i - Y_i] \\
	&= Var[ C_i] + Var[ Y_i] - 2 Cov [C_i, Y_i] \\
	&= 2(\sigmasq + \sigmasqmu) - 2 \sigmasqmu = 2 \sigmasq,
\end{align*}
and
\begin{align*}
E[(\bigcbar - \bigybar)^2] &= Var [\bigcbar - \bigybar] \\
	&= Var [\bigcbar ]+ Var [\bigybar ]- 2 Cov [\bigcbar, \bigybar] \\
	&= \frac{2}{n} (\sigmasq + \sigmasqmu) - \frac{2}{n} \sigmasqmu = \frac{2 \sigmasq}{n}.
\end{align*}
Hence,
\begin{align*}
E [S] = \frac{1}{2a} (2n\sigmasq - 2\sigmasq) = \frac{n-1}{a} \sigmasq.
\end{align*}
The MSE of the estimator is then
\begin{align*}
E [(S - \sigmasq)^2] &= Var [S] + (E [S] - \sigmasq)^2 \\
	&= \frac{1}{2a^2} \bigg( 4n \sigma^4 + \frac{1}{n} \bigg( (6-n)\sigma^4 + (4-2n) \sigmasq \sigmasqmu + \sigma^4_\mu \bigg) - 8 \sigma^4 \bigg) \\
	& \;\;\;\; + \bigg( \frac{n-1}{a} - 1 \bigg)^2 \sigma^4 \\
	&= \frac{1}{2a^2} \bigg( 4n \sigma^4 + \frac{1}{n} \bigg( (6-n)\sigma^4 + (4-2n) \sigmasq \sigmasqmu + \sigma^4_\mu \bigg) - 8 \sigma^4 + 2(n-1)^2 \sigma^4 \bigg) \\
	& \;\;\;\; - 2(n-1)\sigma^4 \frac{1}{a} + \sigma^4 \\
	&= \frac{1}{2a^2} \bigg( (2n^2 + \frac{6}{n} - 7)\sigma^4 + 2(\frac{2}{n}-1)\sigmasq \sigmasqmu + \frac{1}{n} \sigma^4_\mu \bigg)  - 2(n-1)\sigma^4 \frac{1}{a} + \sigma^4. 
\end{align*}
The value of $a$ that minimizes this MSE is
\begin{align*}
a &= \frac{(2n^3 - 7n + 6) \sigma^4 + 2(2-n)\sigmasq \sigmasqmu + \sigma^4_\mu}{2(n^2 - n) \sigma^4} \\
	&= \frac{2n^3 - 7n + 6}{2(n^2 - n)} + \frac{2-n}{n^2-n} \frac{\sigmasqmu}{\sigmasq} + \frac{1}{2(n^2-n)} \bigg( \frac{\sigmasqmu}{\sigmasq} \bigg)^2.
\end{align*}

\section{Calculating $Var [\sinttilde]$}
\label{sec:app.var.int}
\begin{align*}
Var [\sinttilde] &= \frac{1}{4a^2} Var \bigg[ \sumi (C_i - Y_i)^2 \bigg] \\
		 &= \frac{1}{4a^2} Var \bigg[ \sumi \bigg(C_i^2 + Y_i^2 - 2C_i Y_i \bigg) \bigg] \\
		 &= \frac{1}{4a^2} Var \bigg[ \sumi C_i^2 + \sumi Y_i^2 - 2 \sumi C_i, Y_i \bigg]  \\
		 &= \frac{1}{4a^2} \bigg( Var \left[\sumi C_i^2\right] + Var \left[\sumi Y_i^2\right] + 4 Var\left[ \sumi C_i Y_i\right] + 2Cov \left[\sumi C_i^2, \sumi Y_i^2\right] \\
		 & \;\;\;\; - 4 Cov \left[\sumi C_i^2, \sumi C_i Y_i\right] - 4 Cov \left[\sumi Y_i^2, \sumi C_i Y_i\right] \bigg).
\end{align*}
The individual terms can be computed as follows:
\begin{align*}
Var \left[ \sumi C_i^2 \right] &= \sumi Var [C_i^2] \\
				 &= \sumi \bigg(E[C_i^4] - (E [C_i^2])^2 \bigg) \\
				 &= \sumi \bigg( E[C_i^4] - (Var[ C_i] + (E[C_i])^2)^2 \bigg) \\
				 &= \sumi \bigg( E[C_i^4] - (\sigmasq + \sigmasqmu + \mu^2)^2 \bigg) \\
				 &= n EC_1^4 - n(\sigmasq + \sigmasqmu + \mu^2)^2.
\end{align*}
Assuming normality, we have
\begin{align*}
Var \left[ \sumi C_i^2 \right] &= n\bigg( 3\epsilon + 3\sigma^4 + 6\sigmasqmu \sigmasq + 6\mu^2 \sigmasq + 3\sigma^4_\mu + 6\mu^2 \sigma^2_\mu + \mu^4 - (\sigmasq + \sigmasqmu + \mu^2)^2 \bigg) \\
		&= n(3\epsilon + 2\sigma^4 + 2\sigma_\mu^4 + 4\sigmasq \sigmasqmu + 4\mu^2 \sigmasq + 4\mu^2 \sigmasqmu).
\end{align*}
Assuming additionally that $\mu=0$ and $\epsilon=0$, we have
\begin{align*}
Var \left[ \sumi C_i^2\right] = 2n(\sigmasq + \sigmasqmu)^2.
\end{align*}
Since $C_i$ and $Y_i$ are symmetrically defined, we have
\begin{align*}
Var \left[ \sumi Y_i^2 \right] = Var \left[\sumi C_i^2 \right]. 
\end{align*}
Next,
\begin{align*}
Var \left[ \sumi C_i Y_i \right] &= \sumi Var [C_i Y_i] \\
				    &= \sumi \bigg( E[C_i^2 Y_i^2] - (E[C_i Y_i])^2  \bigg)
\end{align*}
where 
\begin{align*}
E[C_i Y_i]^2 &= E\bigg[ E[ C_i^2 Y_i^2 | Z_i \bigg] ] \\
			   &= E[ E[C_i^2 | Z_i) E(Y_i^2 | Z_i]  ] \\
			   &= E[\Sigma_i^2 + M_i^2]^2 \\
			   &= E[\Sigma_i^4 + M_i^4 + 2\Sigma_i^2 M_i^2] \\
			   &= Var [\Sigma_i^2 ]+ (E[\Sigma_i^2])^2 + E[M^4_i] + 2 E[\Sigma_i^2 ]E[M_i^2]  \\
			   &= \epsilon + \sigma^4 + E[M^4_i]+ 2 \sigma^2 (\sigmasqmu + \mu^2);
\end{align*}
and
\begin{align*}
E[C_i Y_i] &= Cov [C_i, Y_i]+ E[C_i] E[Y_i] \\
		      &= \sigmasqmu + \mu^2.
\end{align*}
Therefore,
\begin{align*}
Var \left[ \sumi C_i Y_i \right] &= \sumi \bigg( \epsilon + \sigma^4 + EM^4_i + 2 \sigma^2 (\sigmasqmu + \mu^2) -  (\sigmasqmu + \mu^2)^2 \bigg).
\end{align*}
Assuming normality, we have
\begin{align*}
E[C_i Y_i]^2 &= \epsilon + \sigma^4 + 3\sigma^4_\mu + 6\mu^2 \sigma^2_\mu + \mu^4 + 2\sigmasq \sigmasqmu + 2\sigmasq \mu^2; \\
E[C_i Y_i] &= \sigmasqmu + \mu^2; \\
Var\left[ \sumi C_i Y_i \right] &= n (\epsilon + \sigma^4 + 2\sigma^4_\mu + 2\sigmasq \sigmasqmu + 2\mu^2 \sigmasq + 4\mu^2 \sigma^2_\mu).
\end{align*}
Assuming additionally that $\mu=0$ and $\epsilon=0$, we have
\begin{align*}
E[C_i Y_i]^2 &= (\sigmasq + \sigmasqmu)^2 + 2\sigma_\mu^4; \\
E[C_i Y_i] &= \sigmasqmu; \\
Var \left[ \sumi C_i Y_i \right] &= n\bigg[ (\sigmasq + \sigmasqmu)^2 + \sigma_\mu^4 \bigg].
\end{align*}

The covariance terms are computed as follows:
\begin{align*}
Cov \left[\sumi C_i^2, \sumi Y_i^2\right] =  \sumi Cov [C_i^2, Y_i^2] = \sumi (E[C_i^2 Y_i^2 ] - E[C_i^2 ]E[Y_i^2]).
\end{align*}
Assuming normality, we have
\begin{align*}
Cov \left[\sumi C_i^2, \sumi Y_i^2\right] &= n\bigg( \epsilon + \sigma^4 + 3\sigma^4_\mu + 6\mu^2 \sigma^2_\mu + \mu^4 + 2\sigmasq \sigmasqmu + 2\sigmasq \mu^2 - (\sigmasq + \sigmasqmu + \mu^2)^2 \bigg) \\
	&= n(\epsilon + 2\sigma_\mu^4 + 4\mu^2 \sigmasqmu).
\end{align*}
Assuming additionally that $\mu=0$ and $\epsilon=0$, we have
\begin{align*}
Cov \left[\sumi C_i^2, \sumi Y_i^2\right] = 2n\sigma_\mu^4.
\end{align*}

Finally, since $C_i$ and $Y_i$ are symmetrically defined, we have
\begin{align*}
Cov \left[\sumi C_i^2, \sumi C_i Y_i\right] &= Cov \left[\sumi Y_i^2, \sumi C_i Y_i\right] \\
						   &= \sumi Cov [C_i^2, C_i Y_i] \\
						   &= \sumi \bigg( E[C^3_i Y_i] - E[C^2_i] E[C_i Y_i] \bigg),
\end{align*}
where
\begin{align*}
E[C^3_i Y_i] = E\bigg[ E[C^3_i Y_i | Z_i]  \bigg] = E\bigg[ E[C^3_i | Z_i] E[Y_i | Z_i] \bigg].
\end{align*}
Assuming normality, we have
\begin{align*}
E[C^3_i Y_i] &= E\bigg[ (3M_i \Sigma^2_i + M^3_i) M_i \bigg]  \\
		   &= E[3M^2_i \Sigma^2_i + M^4_i]\\
		   &= 3 E[M^2_i] E[\Sigma^2_i] + E[M^4_i ]\\
		   &= 3 (\sigmasqmu + \mu^2) \sigmasq + 3\sigma^4_\mu + 6\mu^2 \sigma^2_\mu + \mu^4 \\
		   &= \mu^4 + 3\sigma_\mu^4 + 3\sigmasq \sigmasqmu + 3\mu^2 \sigmasq + 6\mu^2\sigmasqmu; \\
E[C^2_i] &= \sigmasq + \sigmasqmu + \mu^2; \\
E[C_i Y_i ]&= \sigmasqmu + \mu^2; 
\end{align*}
and therefore,
\begin{align*}
Cov \left[\sumi C_i^2, \sumi C_i Y_i\right]&= n\bigg(\mu^4 + 3\sigma_\mu^4 + 3\sigmasq \sigmasqmu + 3\mu^2 \sigmasq + 6\mu^2\sigmasqmu - (\sigmasq + \sigmasqmu + \mu^2)(\sigmasqmu + \mu^2) \bigg) \\
	&= n\bigg(\mu^4 + 3\sigma_\mu^4 + 3\sigmasq \sigmasqmu + 3\mu^2 \sigmasq + 6\mu^2\sigmasqmu - (\mu^4 + \sigma_\mu^4 + \sigmasq \sigmasqmu + \mu^2 \sigmasq + 2\mu^2 \sigmasqmu) \bigg)  \\
	&= 2n(\sigma_\mu^4 + \sigmasq \sigmasqmu + \mu^2 \sigmasq + 2\mu^2 \sigmasqmu).
\end{align*}
Assuming additionally that $\mu=0$ and $\epsilon=0$, we have
\begin{align*}
E[C^3_i Y_i] &= 3\sigmasq \sigmasqmu + 3\sigma_\mu^4; \\
E[C^2_i] &= \sigmasq + \sigmasqmu; \\
E[C_i Y_i ]&= \sigmasqmu; \\
Cov \left[\sumi C_i^2, \sumi C_i Y_i\right] &= 2n\sigmasqmu (\sigmasq + \sigmasqmu).
\end{align*}
Putting the terms together, 
we derive the variance as follows, assuming that $M_i$ follows a normal distribution, 
\begin{align*}
Var [\sinttilde] &= \frac{1}{4a^2} \bigg\{ 2n(3\epsilon + 2\sigma^4 + 2\sigma_\mu^4 + 4\sigmasq \sigmasqmu + 4\mu^2 \sigmasq + 4\mu^2 \sigmasqmu) \\
			& \;\;\;\; + 4n (\epsilon + \sigma^4 + 2\sigma^4_\mu + 2\sigmasq \sigmasqmu + 2\mu^2 \sigmasq + 4\mu^2 \sigma^2_\mu) + 2n(\epsilon + 2\sigma_\mu^4 + 4\mu^2 \sigmasqmu) \\
			& \;\;\;\; - 16n(\sigma_\mu^4 + \sigmasq \sigmasqmu + \mu^2 \sigmasq + 2\mu^2 \sigmasqmu)\bigg\} \\
			&= \frac{n}{a^2} (3\epsilon + 2\sigma^4).
\end{align*}
Assuming additionally that $\mu=0$ and $\epsilon=0$, we have
\begin{align*}
Var[ \sinttilde] = \frac{2n}{a^2}\sigma^4.
\end{align*}

\section{Calculating $Var [\sext]$}
\label{sec:app.var.ext}
\begin{align*}
Var [\sext] &= Var \bigg[ \frac{1}{a} (\sumi C_i Y_i - n \bigcbar \bigybar) \bigg] \\
	       &= \frac{1}{a^2} Var \bigg[\sumi C_i Y_i - n \bigcbar \bigybar \bigg] \\
	       &= \frac{1}{a^2} \bigg( Var\left[ \sumi C_i Y_i \right]+ Var [n \bigcbar \bigybar] - 2 Cov \bigg[\sumi C_i Y_i, n \bigcbar \bigybar \bigg]  \bigg). \\
\end{align*}
Here,
\begin{align*}
Var \left[ \sumi C_i Y_i \right] &= \sumi \bigg( \epsilon + \sigma^4 + E[M^4_i ]+ 2 \sigma^2 (\sigmasqmu + \mu^2) -  (\sigmasqmu + \mu^2)^2 \bigg).
\end{align*}
Also,
\begin{align*}
Var [n \bigcbar \bigybar] &= n^2 Var \bigg[ \frac{C_1 + \cdots + C_n}{n} \cdot \frac{Y_1 + \cdots + Y_n}{n}\bigg] \\
				       &= \frac{n^2}{n^4} Var \bigg[\sum_k C_k Y_k + \sum_{i\neq j} C_i Y_j \bigg] \\
				       &= \frac{1}{n^2} \bigg( Var  \left[ \sum_k C_k Y_k \right] + Var \left[ \sum_{i\neq j} C_i Y_j  \right]+ 2Cov \bigg[ \sum_k C_k Y_k, \sum_{i\neq j} C_i Y_j \bigg] \bigg).
\end{align*}
Assuming normality on $M_i$ and assuming that $\mu=0$ and $\epsilon=0$ (constant variance across cells), we have
\begin{align*}
Var \left[ \sum_k C_k Y_k \right] &= n(\sigma^4 + 3\sigma_\mu^4 + 2\sigmasq \sigmasqmu - \sigma_\mu^4) \\
						  &= n(\sigmasq + \sigmasqmu)^2 + n\sigma_\mu^4.
\end{align*}
Also,
\begin{align*}
Var \left[ \sum_{i\neq j} C_i Y_j \right] &= \sum_{i\neq j} Var [C_i Y_j ]+ 2\sum_{i=k\; \text{or} \; j=l} Cov [C_i Y_j, C_k Y_l] + 2\sum_{i\neq k \; \text{and} \; j \neq l} Cov [C_i Y_j, C_k Y_l].
\end{align*}
Under the assumptions made above, we have
\begin{align*}
Var[ C_i Y_j] &= E[C_i^2 Y_j^2] - (E[C_i Y_j])^2 \\
			   &= E[C_i^2] E[Y_j^2] - (E[C_i] E[Y_j])^2 \\
			   &= (\sigmasq + \sigmasqmu)^2.
\end{align*}
If $i=k$,
\begin{align*}
Cov [C_i Y_j, C_k Y_l] &= E[C_i Y_j C_k Y_l ]- E[C_i Y_j ]E[C_k Y_l] \\
							&= E[C_i^2] E[Y_j ]E[Y_l] - (E[C_i])^2 E[Y_j] E[Y_l] \\
							&= 0.
\end{align*}
Similarly, we can derive that the covariance is 0 for other cases where $j=l$ or where $i\neq k$ and $j\neq l$.  Hence,
\begin{align*}
Var \left[ \sum_{i\neq j} C_i Y_j \right] = n(n-1)(\sigmasq + \sigmasqmu)^2.
\end{align*}
Additionally, under the normality assumption and with $\mu=0$ and $\epsilon=0$,
\begin{align*}
Cov \bigg[ \sum_k C_k Y_k, \sum_{i\neq j} C_i Y_j \bigg] = 0.
\end{align*}
Therefore,
\begin{align*}
Var [n \bigcbar \bigybar] &= \frac{1}{n^2} \bigg( n(\sigmasq + \sigmasqmu)^2 + n\sigma_\mu^4 + n(n-1)(\sigmasq + \sigmasqmu)^2 \bigg) \\
							   &=  \frac{1}{n^2} \bigg( n^2 (\sigma^2 + \sigma_\mu^2)^2 + n\sigma_\mu^4 \bigg) \\
							   &= (\sigma^2 + \sigma_\mu^2)^2 + \frac{\sigma_\mu^4}{n}.
\end{align*}
Furthermore,
\begin{align*}
Cov \bigg[\sumi C_i Y_i, n \bigcbar \bigybar \bigg] &= \frac{1}{n} Cov \left[\sumi C_i Y_i, \sum_k C_k Y_k + \sum_{i\neq j} C_i Y_j\right]  \\
	&= \frac{1}{n} \bigg( Cov \bigg[\sumi C_i Y_i, \sum_k C_k Y_k\bigg] + Cov \bigg[\sumi C_i Y_i, \sum_{i\neq j} C_i Y_j\bigg] \bigg) \\
	&= \frac{1}{n} \bigg( Var \left[ \sumi C_i Y_i \right]  \bigg) \\
	&= (\sigmasq + \sigmasqmu)^2 + \sigma_\mu^4.
\end{align*}
\begin{align*}
Var[ \sext] &= \frac{1}{a^2} \bigg( n(\sigmasq + \sigmasqmu)^2 + n\sigma_\mu^4 + (\sigma^2 + \sigma_\mu^2)^2 + \frac{\sigma_\mu^4}{n} - 2(\sigmasq + \sigmasqmu)^2 - 2\sigma_\mu^4 \bigg) \\
		    &= \frac{n-1}{a^2} (\sigma^2 + \sigma_\mu^2)^2 + \frac{(n-1)^2}{na^2} \sigma_\mu^4.
\end{align*}

\end{document}